\documentclass[12pt,a4paper]{article}
\usepackage[a4paper, total={6in,9in}]{geometry}
\usepackage{amsmath}
\usepackage{amsfonts}
\usepackage{amssymb}
\usepackage{graphicx}
\usepackage{color}
\usepackage{epstopdf}
\usepackage{xcolor,colortbl}
\usepackage{caption}
\usepackage{tikz}
\usepackage{subfigure}

\newcommand*{\sectioncolor}{black}
\newcommand*{\sectionformat}{\centering\color{\sectioncolor}}

\usepackage{ulem}
 
\begin{document}

\begin{center}
\begin{large}
\textbf{
Nanotribology of ionic liquids: transition to yielding response in nanometric confinement with metallic surfaces  
}
\end{large}
\end{center}

\begin{center}
Antoine Lain\'e, Antoine Nigu\`es, Lyd\'eric Bocquet and Alessandro Siria$^{*}$\\
\vspace{0.2cm}
\textit{Laboratoire de Physique de l'\'Ecole Normale Sup\'erieure, ENS, Universit\'e PSL, CNRS, Sorbonne Universit\'e, Universit\'e Paris-Diderot, Sorbonne Paris Cit\'e, UMR CNRS 8550, 24 Rue Lhomond 75005 Paris, France}\\
$^*$ alessandro.siria@lps.ens.fr
\end{center}

{\bf
Room Temperature Ionic Liquids (RTILs) are molten salts which exhibit uniques physical and chemical properties, commonly harnessed for lubrication and energy applications. The pure ionic nature of RTIL leads  to strong electrostatic interactions among the liquid, furthermore exalted in the presence of interfaces and confinement. In this work, we use a tuning-fork based dynamic Surface Force Tribometer (TF-SFT), which allows probing both the rheological and the tribological properties of RTILs films confined between a millimetric sphere and a surface, over a wide range of confinements. When the RTIL is confined between metallic surfaces, we evidence an abrupt change of its rheological properties below a threshold confinement. This is reminiscent of a recently reported confinement induced capillary freezing, here observed with a wide contact area. 
In parallel, we probe the tribological response of the film under imposed nanometric shear deformation and unveil a yielding behaviour of the interfacial solid phase below this threshold confinement. This is characterized by a transition from an elastic to a plastic regime, exhibiting striking similarities with the response of glassy materials. This transition to yielding  of the RTIL in metallic confinement leads overall to a reduction in friction and offers a self-healing protection of the surfaces avoiding direct contact, with obvious applications in tribology. 

}

\clearpage
\section*{\sectionformat I. Introduction}
\vskip0.2cm
 Room Temperature Ionic liquids (RTILs) are intriguing materials: they exhibit  standard features of liquid matter, but their molten salt nature, associated the strong coulombic interaction between molecules, is quite unique and leads to a variety of specific properties which are harnessed in numerous of applications from electro-chemistry to tribology. Their behavior usually challenges well established frameworks inherited from liquid state theory or that of dilute electrolytes\cite{Smith2016,Gebbie2017,Rotenberg2018}. This unique response is furthermore exalted under confinement or at interfaces, and a rich variety of behaviors have been been observed for RTILs under confinement, \cite{Maali2008,Perkin2012,Perkin2011,Futamura2017,Galluzzi2018} with a strong dependence on the nature of the interfaces \cite{Jurado2017,Lee2016,Voeltzel2018,Li2013}.
Experimentally, Atomic Force Microscopy (AFM) \cite{Hoth2014} is a technique of choice to probe the properties of RTIL in confinement, as well as Surface Force Apparatus (SFA) \cite{Tomita2018,Smith2013}. A notable difference between the two approaches is the lateral size of the probed contact, which ranges from tens of nanometer for AFM to microns in SFA. Also properties are mostly investigated with insulating confining surfaces, but recent work has put forward the specificity of metallic confinement for RTILs, with the evidence of a confinement induced capillary freezing resulting associated with the metallic nature of the surfaces \cite{Comtet2017}. That the nature of electric boundary conditions may affect the properties of charged molten salt is actually quite expected \cite{Kaiser2017}. However, the impact of metallic confinement on the properties of confined RTILs remains to be fully assessed and calls for new experimental investigation.  Furthermore, metallic surfaces are ubiquitous in applications of RTILs, {\it e.g.} for lubrication, where RTILs are expected to be excellent boundary lubricants \cite{Lhermerout2018,Gong2018}, as well as in electrochemistry \cite{Armand2009}. 
In the present work, we use a newly introduced force measurement methodology, based on a macroscopic Tuning Fork \cite{Canale2018,Laine2019}  and here combined with electric measurements, in order to probe the equilibrium and out-of-equilibrium frictional properties of RTILs confined between extended metallic surfaces. This instrument allows to probe both the rheology of the confined materials, as well as its tribological properties under shear.  As in Ref. \cite{Comtet2017}, the ionic liquid under scrutinity is BmimBF$_4$, exhibiting a viscosity $\eta \approx 50 - 130$ mPa.s at room temperature \cite{Tomida2006} and a bulk freezing temperature $T_B \approx -71 ^{\circ}$C \cite{Comtet2017}.

\begin{figure}[htb!]
\centering
\includegraphics[width=0.94\columnwidth]{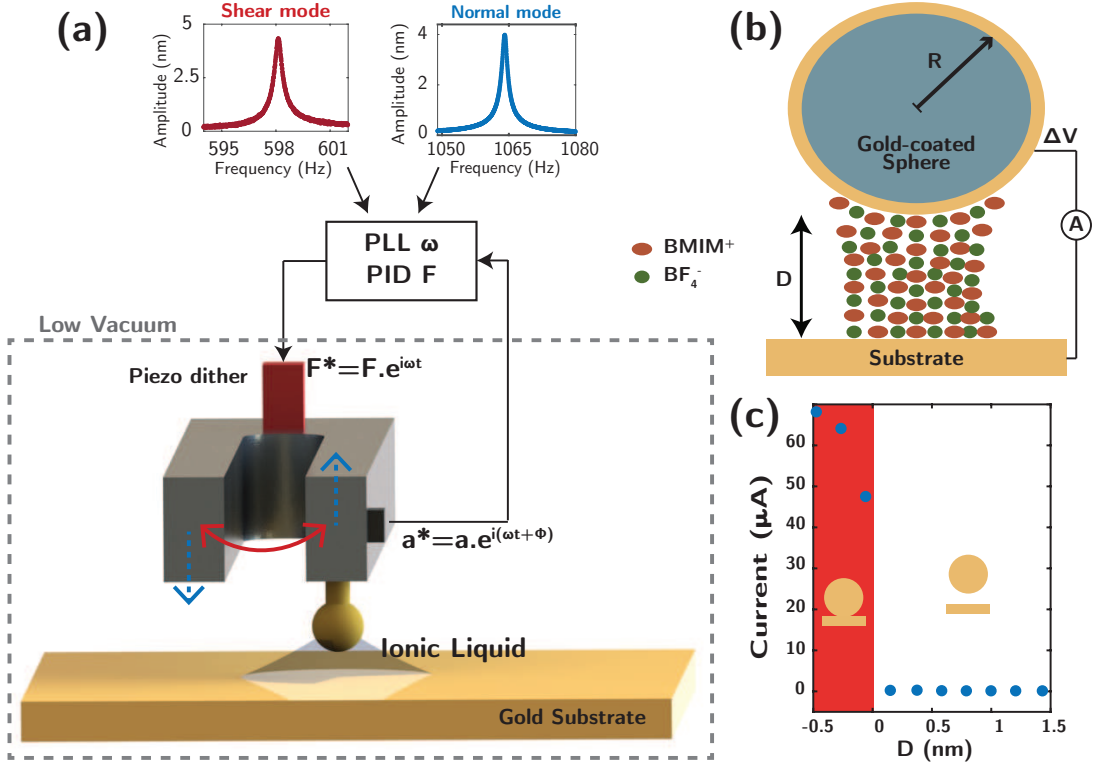}
\caption{\textbf{Experimental setup} \textbf{(a)} Schematic representation of the force measurement apparatus. The prongs of the tuning fork oscillate along a shear (red) and a normal (blue) mode simultaneously. The two curves at the top show the corresponding resonance curves. The mechanical excitation $F^* = F \cdot e^{i \omega t}$ [N] is induced by the actuation of an external piezodither. An accelerometer glued on one prong records the dynamic oscillation amplitude $a^* = a \cdot e^{(i \omega t + \phi)}$ [m]. A lock-in amplifier extracts the amplitude and phase of the accelerometer signal. The phase difference is continuously kept at zero by use of a Phase-Locked Loop (PLL) which varies the excitation frequency ($\omega$) accordingly. A feedback loop (PID) controls the excitation intensity $F$ in order to keep the oscillation amplitude a chosen value. A spherical probe, glued at the end of one prong, is immersed in an Ionic Liquid drop deposited on a gold substrate. The set-up is placed in a reduced pressure ($P<10^{-1}$ mbar) environment. \textbf{(b)} The Ionic Liquid BmimBF$_4$ is confined between the substrate and the gold-coated sphere of radius $R \approx 1.5$ mm. A piezoscanner varies the confinement distance $D$ over tens of micrometers with subnanometric precision. A potential difference $\Delta V \approx 25$mV is applied between the polarizable surfaces. \textbf{(c)} Electric current flowing through the contact upon decreasing the separation distance $D$.}
\label{fig:exp-setup}
\end{figure}

\section*{\sectionformat II. Experimental setup and materials}

A gold-coated sphere of radius $R \sim 1.5$ mm is glued on one prong of a centimetric aluminium tuning fork (Figure \ref{fig:exp-setup} \textbf{(a)}) used as a force sensor. A piezo dither mechanically excites the tuning fork and an accelerometer monitors the oscillation amplitude and phase shift along the normal (blue) and shear (red) directions. In the following we will refer to the {\it normal probe} as the nanorheology probe since it provides information on the visco-elastic properties of the confined fluid; conversely we will refer to the {\it shear mode} as the nanotribology probe since it provides information on the mechanical shear response under an imposed excitation. The present instrument allows probing the two modes simultaneously. \\
A piezoscanner imposes the confinement distance $D$ over several micrometers with sub-nanometer precision. The two polarizable surfaces enable to impose a voltage difference $\Delta V$ and record the electric current flowing through the contact (Figure \ref{fig:exp-setup} \textbf{(b)}) with an I/V amplifier. To prevent the contamination of RTILs from ambient humidity and possible modification of the mechanical properties of nanoconfined RTILs \cite{Jurado2016}, the entire setup is kept in a vacuum chamber with the pressure below $10^{-1}$ mbar .
The centimetric tuning fork serves as force sensor and enables simultaneous normal and shear measurements by exciting two orthogonal modes of the tuning fork \cite{Laine2019}, both presenting a high stiffness $K \sim 10^5-10^6$ N/m and a high quality factor $Q_0 \sim 10^3$. The oscillation amplitude range lies from 200 nm down to 2 nm. As in standard frequency modulation AFM \cite{Canale2018} a conservative interaction leads to a shift of the oscillator resonance frequency $f_0 \rightarrow f_0 + \delta f $ and a dissipative force implies an extra damping which broadens the resonance $Q_0 \rightarrow Q_0 - \delta Q$. Despite the size of the force probe, this experimental set-up was already shown to display the force sensitivity required for atomic force microscopy \cite{Canale2018} and even detection of subnanometric layering of RTILs confined between a (insulating) mica surface and glass bead\cite{Laine2019}.\\
A voltage difference of $\Delta V \approx 25$ mV is furthermore applied between the  gold surface and the gold-coated sphere  (figure \ref{fig:exp-setup} \textbf{(b)}). This allows us to determine the absolute zero separation distance from the electric current response  (figure \ref{fig:exp-setup} \textbf{(c)}). Indeed, the  electronic conductivity of RTILs 
is negligible \cite{Galluzzi2018}, so that the strong increase of the current as observed in Fig. \ref{fig:exp-setup} \textbf{(c)} is the signature of the direct contact between the gold  surface with the gold-coated sphere.

\begin{figure}[h]
\centering
\includegraphics[width=\columnwidth]{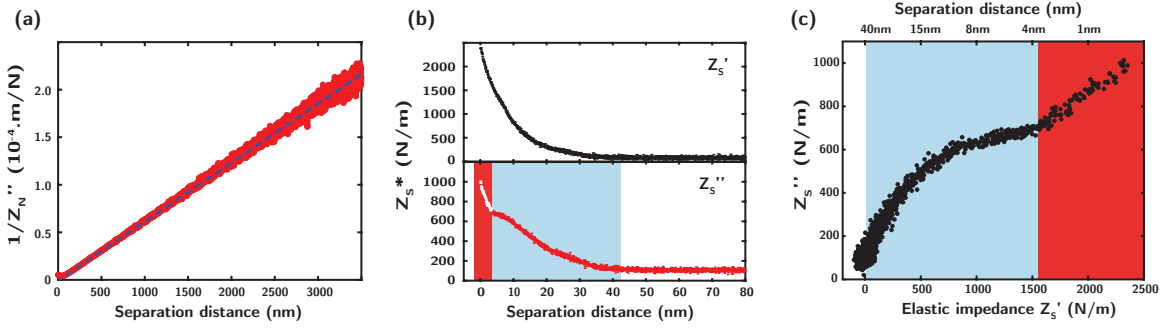}
\caption{\textbf{Nanorheological and nanotribological measurements} \textbf{(a)} Inverse of the normal dissipative impedance $1/Z_N''$ as a function of separation between the surfaces. The dashed line results from a linear fit characteristic of viscous damping. The normal oscillation amplitude is $a_N \sim 2$ nm.  \textbf{(b)} Evolution of the conservative $Z_S'$ (black) and dissipative $Z_S''$ (red) shear impedance as the separation distance is varied. A solid-like mechanical behaviour of the nanoconfined RTILs is observed from $\sim 40$ nm confinement (blue zone) down to the direct metallic contact (red zone and white data points). The shear oscillation amplitude is $a_S \sim 3$ nm. \textbf{(c)} Corresponding shear dissipative impedance $Z_S''$ as a function of shear elastic impedance $Z_S'$ highlighting an abrupt change of frictional behaviour.}
\label{fig:tribo} 
\end{figure}

\section*{\sectionformat III. Nanorheology measurements}

Figure \ref{fig:tribo} presents the simultaneaous measurement of the normal and tangential impedance as the separation distance is decreased from micrometers down to direct contact between the confining metalic surfaces. The important stiffness of the entire setup enables to fully squeeze out the confined liquid until direct surfaces contact is reached. The normal oscillation of the sphere creates a squeeze flow with typical extension $\sim \sqrt{2 R D}$ \cite{Leroy2011}. The associated dissipative normal impedance due to the liquid viscosity thus reads 
\begin{equation}
Z_N'' = \frac{6 \pi \eta R^2 \omega_N}{D-D_0}
\end{equation} 
where $R$ is the sphere radius, $\omega_N = 2\pi f_N$ the normal oscillation frequency, $\eta$ the fluid viscosity and $D_0$ the position of the no-slip plane, usually called hydrodynamic zero. The linear behaviour of the inverse of the normal damping impedance $1/Z_N''$ with respect to separation distance D (figure \ref{fig:tribo} \textbf{(a)}) evidences the liquid-like rheological response down to hundreds of nanometers. The linear fit (dashed blue line) provides a viscosity $\eta \approx 60$ mPa.s comparable with previous measurements \cite{Tomida2006}. Besides, the long range viscous response allows for a determination of the flow boundary condition and provides a hydrodynamic zero $D_0 \approx 10 $ nm with respect to the contact defined by electrical measurements. Assuming that the strong electrostatic interactions between ions and the metallic surfaces prevent any slip at the interface, such non zero $D_0$ means that some solid prewetting films are not fully mobilized by the oscillatory squeeze flow. Such interfacial nanostructures of a few nanometers were evidenced in situ by Atomic Force Microscopy \cite{Comtet2017} as well as Scanning Tunneling Microscopy \cite{Endres2012}. Here, long-range hydrodynamic flow measurement, with a hydrodynamic probe of effective radius $\sqrt{2 R D} \approx 50 ~ \mu$m, accurately resolves the nanometric interfacial features.

\section*{\sectionformat IV. Frictional response}

Simultaneous shear measurements evidence that the shear mechanical impedance $Z_S^*$ of the bulk RTILs remains negligibly small until  nanoconfinement is reached (figure \ref{fig:tribo} \textbf{(b)}). For separation distance  $\approx 50$ nm we observe an important increase of both the elastic (black) and dissipative (red) shear modulus. Further decreasing the separation distance $D$ (blue zone), the shear impedance keeps increasing, as well as the imposed shear strain $\gamma=a_S/D$ and the indentation contact area ${\cal A}_{\rm eff}$, until direct metallic contact is reached (red zone). As we indent into the  interfacial phase (blue zone), the elastic response strongly dominates while the dissipative part tends to saturate around $Z_S'' \approx 0.7$ kN/m corresponding to a dissipation $F_S \approx 2.5 ~ \mu$N. An important conclusion from this measurement is that the predominance of the elastic part highlights the solid-like behavior of the interfacial film, along with the limited frictional dissipation. We emphasize that in the case of standard silicon oil, no such response is obtained (see Supplementary Information, Figure 2). 
Also, an important remark is that this behavior is not observed for the RTIL confined between insulating surfaces (see Supplementary Information, Figure 3).
Here, this behaviour was observed for several approach-retracts at the same and at different locations on the substrate. We report the observation of this shear mechanical response at an average distance $\lambda_S \approx 60$ nm from the electric contact, and  independent on the tangential and normal oscillation amplitude. We note that the measurement of a  critical confinement below which a solid-like behavior is observed is in agreement with the previous report in Ref. \cite{Comtet2017}, although with a probe radius around 3 orders of magnitude bigger. \\
Considering the frictional properties of the RTILs, we plot the evolution of $Z_S''$ as a function of $Z_S'$ (Fig. \ref{fig:tribo} \textbf{(c)}) and we observe a decrease of the slope while dissipating in the solid-like structure (blue zone);  then, above a threshold associated with an elastic interaction $Z_S' \approx 1$ kN/m,  the dissipative impedance tends to saturate with increasing elastic impedance. Finally, for a separation distance under 4 nm, we observe an abrupt change of behaviour and a strong increase of the dissipation (red zone). As the load increases with decreasing D, the abovementioned slope can be qualitatively related to a friction coefficient. Thus, the solid-like nature of the interfacial film of RTIL not only  protects the metallic surfaces from direct contact, but moreover enables to lubricate the contact by reducing the frictional dissipation in the saturated region (where this effective friction coefficient is substantially reduced, see below).

\section*{\sectionformat V. Shear mechanical properties}
\begin{figure}[htb!]
\centering
\includegraphics[width=\columnwidth]{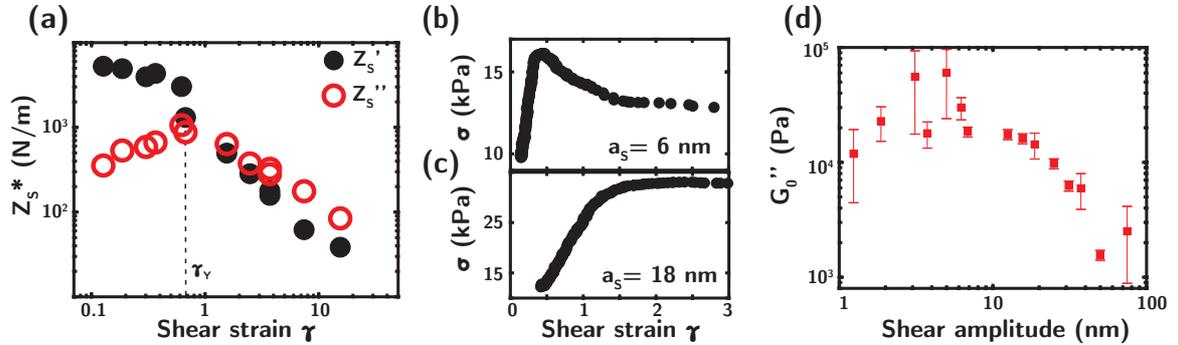}
\caption{\textbf{Shear mechanical properties of the solid interfacial phase} \textbf{(a)} Shear mechanical impedance $Z_S^*$ as a function of applied shear strain for a fixed confinement distance $D=10$ nm. \textbf{(b)-(c)} Stress-strain curves at constant shear amplitude $a_S=6$ nm and $a_S=18$ nm respectively. \textbf{(d)} Dissipative shear modulus $G''$ as a function of shear amplitude obtained from fitting the linear part of the different stress-strain curves. Error bars represent the standard error.}
\label{fig:mech} 
\end{figure}

While we first concentrated on the normal response and the resulting nanorheology properties, we now turn to the investigation of the response of the confined film under shear. This allows us to characterize the frictional properties of the interfacial material under shear by varying the imposed lateral oscillation amplitude. 
We plot in Figure \ref{fig:mech}  \textbf{(a)} the shear mechanical properties of the interfacial solid phase as a function of applied shear strain $\gamma=a_S/D$. In this figure, the confinement is fixed to $D=10$nm.  We observe a  shear softening of the interfacial structure and the elastic and dissipative response do cross above a threshold strain $\gamma_Y$. Such a signature is usually associated with the yielding nature of the investigated material \cite{Mason1996}. The characteristic yield point -- vertical dashed line -- separates an elastic regime at low shear strain and a plastic, dissipation-dominated, regime at high shear strain (figure \ref{fig:mech} \textbf{(a)}).\\
The transition from elastic to plastic mechanical response is also observed from the different stress-strain curves (Fig. \ref{fig:mech} \textbf{(b)-(c)}) where the shear stress is defined here as $\sigma = Z''_S \cdot a_S/{\cal A}_{\rm eff}$  with ${\cal A}_{\rm eff}$ the effective contact area. The solid nature of the interfacial film makes the classical Hertz model of contact mechanics suitable to derive an estimate for the effective contact area. Therefore, we use from now the approximation ${\cal A}_{\rm eff} \approx {\cal A}_{\rm Hertz}$ where ${\cal A}_{\rm Hertz} = 2 \pi R \delta $, with the indentation depth $\delta = \lambda_S - D$. 

We plot in Figure \ref{fig:mech} \textbf{(b)} the stress versus strain curve for two different imposed  shear amplitude $a_S$. For a small shear amplitude, $a_S = 6$ nm (top panel), we observe first a linear regime where $\sigma = \sigma_Y + G'' \cdot \gamma$ before reaching a peak stress  and eventually a plateau at larger strain. Now, Increasing the shear amplitude to $a_S=18$ nm (bottom panel), the stress overshoot disappears and the linear regime is directly followed by a plateau. This qualitative behaviour change as well as the shear dissipative modulus decrease with increasing amplitude (figure \ref{fig:mech} \textbf{(d)}) is reminiscent of the existence of a critical yield point above which the material gets fluidized.
Let us now analyze more in details the various elastic and plastic regimes.

\subsection*{\sectionformat A. Elastic regime}
In the elastic regime, $\gamma<\gamma_Y$, the conservative mechanical impedance is fairly constant and strongly dominates the dissipative part by one order of magnitude (figure \ref{fig:mech} \textbf{(a)}). The imposed shear deformation remains relatively small and the interfacial phase elastically compensates the deformation. From this initial plateau, and using a Hertz contact mechanics model to approximate the contact area ${\cal A}_{\rm eff} \approx {\cal A}_{\rm Hertz} = 2 \pi R (\lambda_S -D) \approx 400~\mu$m$^2$, we obtain an estimate of the elastic shear modulus $G'_0 \approx \frac{D}{{\cal A}_{\rm eff}} Z'_{S0} \approx 130$ kPa. Interestingly, the dissipative shear modulus for small shear amplitude lies around $G''_0 \approx 20$ kPa (figure \ref{fig:mech} \textbf{(d)}) and is therefore one order of magnitude smaller than $G'_0$. For isotropic materials, one gets the relation between shear and Young modulus $E'_0=2G'_0(1+\nu)$, such that we estimate $E'_0 \approx 260$ kPa for a Poisson ratio $\nu =0.2$.

\subsection*{\sectionformat B. Yielding and plastic regime}

By increasing the shear strain above the yield strain, $\gamma>\gamma_Y$, we observe on figure \ref{fig:mech} \textbf{(a)} a transition from an elastic to a plastic regime, where the dissipation overcomes the elastic reponse of the system. Such behaviour is reminiscent of yielding mechanism observed in emulsions and glassy materials \cite{Mason1996,Christopoulou2009}. At the yield point, the elastic impedance sharply decreases and the dissipative impedance reaches its maximum value. From figure \ref{fig:mech} \textbf{(a)} we can extract a value of the yield force $F_Y$, the maximum applied force the system is capable to handle elastically, from the relation $F_Y = Z'_Y \cdot \gamma_Y \cdot D$ where $Z'_Y$ is the elastic impedance and $\gamma_Y$ the shear strain, both at the yield point. We extract the values $Z'_Y \approx 3000$ N/m and $\gamma_Y \approx 0.7$ eventually giving an experimental value of the yield force $F_Y \approx 2.1~\mu$N and a corresponding yield strength $\sigma_Y = F_Y/{\cal A}_{\rm eff} \approx 5$ kPa. 
From fitting the linear part of the stress-strain curves, we estimate $\sigma_Y \approx 4$ kPa in excellent agreement with the first estimate. This value is about one to two orders of magnitude smaller than the estimated Young modulus $E'_0$. This is typical of defect-mediated yielding mechanism and points towards a amorphous or polycrystalline structure of the interfacial film. \\
For shear strain above $\gamma_Y$, the response of the system is dominated by the dissipation. Both the elastic and dissipative parts of the impedance decrease with increasing shear, thus eventually pointing towards a shear induced fluidization of the interfacial structure. The  disappearance of the peak stress  under an increase of the shear amplitude (Fig. \ref{fig:mech} \textbf{(c)}) further corresponds to the mechanical response of an amorphous solid \cite{Jiang2015}. The stress overshoot, possibly resulting from a delayed activation, disappears in the fluidized plastic regime because the activation barrier is more easily crossed.  

\section*{\sectionformat VI. Transition to glassy response in confinement}

The observed mechanical behavior  highlights strong similarities with the mechanical response of glassy materials. In the theoretical interpretation developed in a previous study \cite{Comtet2017,Kaiser2017}, it was argued that the RTIL undergoes a capillary freezing in confinement: the shift of the crystallization temperature results from the stabilization of the solid phase due an image charge effect at the metallic surfaces.  
However the properties of the highly confined material could not be investigated previously: here, using the new instrument, we show that the transition to the solid state under confinement does not reach a state with  crystalline response, but rather exhibit the properties of a yielding, glassy-like state. Such a frustrated transition is to be expected due to kinetic barriers. This is further supported by observations from the literature. Indeed, bulk DSC measurements report that RTILs get supercooled and form glasses rather than crystallize \cite{Fredlake2004}. Besides, 2D IR spectroscopy of RTILs thin films also presents a dynamical molecular behaviour that fully ressembles the one of a supercooled liquid \cite{Nishida2018}.
The mechanical measurements performed on solid-like, fully crystallized, interfacial systems obtained from drop-casting of RTILs/methanol solution provides a measure of the Young modulus $E^* \approx 60$ MPa \cite{Borghi2019}. Here, we obtain a value that is about 2 orders of magnitude smaller and further points towards the glassy nature of the nanoconfined RTILs.

\section*{\sectionformat VII. Shear dissipation-velocity dependence}

\begin{figure}[htb!]
\centering
\includegraphics[width=\columnwidth]{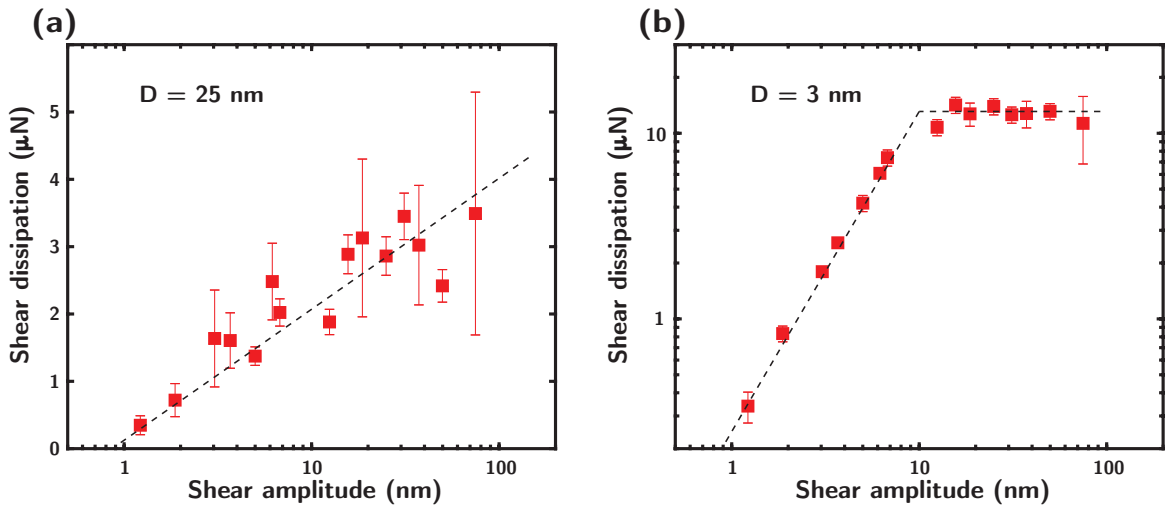}
\caption{\textbf{Shear frictional properties of the solid interfacial phase} \textbf{(a)} Logarithmic dependence of the dissipative friction force with respect to the shear amplitude for a fixed confinement distance $D=25$ nm. \textbf{(b)} Transition from a power-law to solid friction for fixed separation distance $D=3$ nm. Error bars represent the standard error and correspond to 10\% for single points.}
\label{fig:frict} 
\end{figure}

The dependence of the shear dissipation force $F_D$ on the shear velocity $v$ provides further insights into the mechanisms governing the dissipative response of the interfacial glassy structure. As we show on Figure \ref{fig:frict} \textbf{(a)}, for a fixed separation distance $D=25$nm, we observe a logarithmic dependence of $F_D$ with respect to the shear amplitude $a_S$. 
Such response is actually expected for glassy materials undergoing plastic reorganization.
In the framework of a Eyring's theory, commonly used to describe the mechanical response of glassy materials in terms of a shear induced activation process \cite{Varnik2004}, the shear stress reads 
$$\sigma_D=F_D/{\cal A}_{\rm eff}=\frac{k_BT}{V^*}ln\left(\frac{a_S}{a_{S0}^*}\right)$$ 
where $V^*$ the activation volume corresponds to the elementary volume over which a plastic event occurs.
From the fit of the data, we obtain a typical size for the activation volume $l^* \sim \left(V^*\right)^{1/3} \approx 10$ nm. 
 Altogehter this emphasizes the glassy behaviour of the interfacial crystallised layer where the dissipation is triggered by plastic events taking place on a length scale of several nanometers. In this regime, the dissipative events are to occur at the weaker points in the system and these ones are more likely to appear in the bulk than in the close vicinity of the interface where short-range wetting structures are present and strongly pinned to the surface.

Further decreasing the confinement distance down to $D=3$ nm (Fig. \ref{fig:frict} \textbf{(b)}), we observe stricking differences in the frictional response. 
The shear dissipation to amplitude curve exhibits a power-law behaviour $F_{D} \sim a_S^{\alpha}$ at low amplitude with an exponent $\alpha \approx 1.5$, whereas for higher amplitude, above $a_c \approx 10~$nm, solid-like friction is observed with a constant dissipative force. 
Confinement in the latter case is below the characteristic length scale of plasticity, $l^* \approx 10$nm,  obtained above from the elementary activation volume $l^* \sim V^{*\,1/3} $. Accordingly, the different signature of the shear force highlights a change of regime where one probes dissipation below the elementary length scales associated with plastic events.  
This is furthermore consistent with the observation that the transition from the regime in Fig. \ref{fig:frict} \textbf{(a)} to \ref{fig:frict}.\textbf{(b)} appears smoothly around $D \sim 10-15$ nm (not shown).

\section*{\sectionformat VIII. Conclusion}

In this work, we have exploited the capabilities of a recently introduced  tuning-fork based dynamic Surface Force instrument, which allows to perform a joint investigation of nanorheology and nanotribology of confined films with a millimetric probe down to nanoscale confinements. This instrument effectively bridges the gap between atomic force probes and large scale tribology and hence provides an exhaustive overview of the mechanical response of the investigated material in confinement. \\
We  show that a room-temperature ionic liquid (RTIL) confined between {\it metallic surfaces} exhibit a transition towards solid-like mechanical response below a threshold confinement. This observation agrees  with a previous report \cite{Comtet2017}, however highlighted here with a millimetric probe and wide contact area. Furthermore the joint nano-rheology/-tribology approach allowed by our instrument shows that the ionic liquid undergoes a transition towards a glassy state, with a yielding behavior above a threshold shear strain. Previous theoretical work interpreted the change of mechanical response of the ionic liquid as a freezing transition \cite{Comtet2017,Kaiser2017} occuring in metallic confinement due to image charge effects. However the present work demonstrates that the materials does not exhibit the response of a fully crystalline materials, but rather the yielding response of a glassy material. Accordingly this indicates that the predicted freezing transition in confinment is either incomplete (leading to a polycrystalline structure) or frustrated (leading to a amorphous structure).

With the increasing appeal to use RTILs as electrolyte in supercapacitor, our results indicate that some care should be taken while designing such devices involving an important interfacial contact between RTILs and metallic surfaces since the simple picture of a fully liquid electrolyte may not be valid anymore.

More important our observations points to key applications in tribology. Indeed the confinement-induced transition of the ionic liquid transforms the liquid into a solid-like film which acts as a good lubricating layer and reduces frictional dissipation. The interfacial solid phase is self-consitently generated under confinement between metallic surfaces and therefore acts as a `sacrificial' anti-wear film, protecting the surfaces from direct contact. 
Such characteristics, in addition to manifold appealing physical properties of RTILs (low vapor pressure, high thermal stability and wide electrochemical window), makes them ideal candidates as anti-wear boundary lubricants \cite{Gosvami2015}. We emphasize that the response reported here is observed under vacuum conditions and the presence of water in RTILs may screen the electrostatic interactions at stake and then reduce or even mask the mechanical properties highlighted here. Then it may prove to be important to prefer hydrophobic RTILs in order to take fully advantage of such a mechanical response.Furthermore, future  work may also explore the capabilities of RTILs as {\it active lubricant}, via the control of interfacial properties thanks to surfaces electric polarization \cite{Lhermerout2018,Sweeney2012, Dold2015,Fajardo2015}. This will open new routes for unforeseen lubricating mechanisms.

\bibliographystyle{ieeetr}
\bibliography{TIL2}

\end{document}